# Interface and Data Biopolitics in the Age of Hyperconnectivity. Implications for Design

Salvatore Iaconesi[a]

[a]ISIA Design Florence
*Corresponding author e-mail: salvatore.iaconesi@artisopensource.net

**Abstract:** This article describes their biopolitical implications for design from psychological, cultural, legal, functional and aesthetic/perceptive ways, in the framework of Hyperconnectivity: the condition according to which person-to-person, person-to-machine and machine-to-machine communication progressively shift to networked and digital means.
A definition is given for the terms of "interface biopolitics" and "data biopolitics", as well as evidence supporting these definitions and a description of the technological, theoretical and practice-based innovations bringing them into meaningful existence.
Interfaces, algorithms, artificial intelligences of various types, the tendency in quantified self and the concept of "information bubbles" will be examined in terms of interface and data biopolitics, from the point of view of design, and for their implications in terms of freedoms, transparency, justice and accessibility to human rights.
A working hypothesis is described for technologically relevant design practices and education processes, in order to confront with these issues in critical, ethical and inclusive ways.

## 1. A Hymn

In her "Hymn of Acxiom" folk singer Vienna Teng (2013) starts off with lyrics "Somebody hears you, you know that...", in what seems to be a church choir. After listening for a bit, the real topic the artist is discussing about becomes clear: Acxiom is not a benevolent divinity somewhere in the cosmo-sphere caringly waiting to hear the troubles of his beloved human beings, but, rather, a high-powered data broker which has been described as "the Private NSA" (Tom's Guide, 2013), as the silent, largest consumer data processor in the world (Fortune Magazine, 2004) and as "Big Brother in Arkansas" (NY Times, 2012). The topic of the song is data-surveillance. The idea for the song came while the author was pursuing an MBA at the University of Michigan: a colleague working with Acxiom data was shocked about the amount of information the company had available about herself and her husband. An interesting thing about the song is that the creepy, Orwellian, lyrics also empathize with databases as well as excoriating them.

This is, in fact, an interesting point of view. As, on the one hand, we – directly and indirectly – consent tour data to be collected through our behaviors and basically accepting our lifestyle, on the other hand we feel deeply uneasy about that and from its implications. As we benefit from enterprises being able to provide us with products and services which are "more relevant" for us (more on this later in the article), we are simultaneously wary of the fact that a single subject knows so much about us and uses it to "sell us" to the highest bidder, and what this possibility implies for our liberties, freedoms and rights. Even more, it is progressively harder, if not impossible, for us, to know and understand what parts of our online and offline environment are determined from these data collection processes, or about which subjects have this data about us available, or how they are using it and for what purposes (Lafrance, 2014).

In fact, the entries we see when we browse search engines, social media websites, and other online services are completely determined by these processes: the operators of these services feed the data they acquired about us to their classification algorithms which, in return, use it to determine what we may be more and less interested about, or where their optimal business opportunities lay in relation to our profile.

This is not only true for our online lives: the offline world is quickly catching up.  Physical mailings; fidelity cards; algorithmic research and control applied to stores, services and spaces of the city; imaging through security cameras; object and facial recognition on devices and architectures; the Internet of Things (IoT) and its sensors, possibilities for user identification and biometrics. These all combine with a plethora of other options which are turning us into data subjects, which can be recognized and tracked on databases as in the physical world.

## 2. Asymmetry

This scenario also describes a progressive asymmetry in the distribution of power, rights, freedoms and opportunities (Tufekci, 2014; boyd, 2012).

As a matter of fact, it is practically and psychologically impossible for human beings to understand which and how much data is captured about them, how and why it is used, and what effects it has about their lives.

The complex interplay among users; organisation; algorithms; national, international and global regulations and agreements, or lack of them; data and information flows within user experiences in the physical and online domains cause grey areas to emerge, at levels which are legal, cultural, psychological, ethical and philosophical (White, 2016).

"Code is Law", Lawrence Lessig (2006) once said. And this is really the case nowadays. With thousands of updates and modifications to the interfaces, algorithms, data capture and usage profiles which are performed each month to the systems of popular services, potentially provoking radical changes to the implications for privacy, control and accountability, it is practically impossible for legal and cultural systems not only to adapt and react, but also and more importantly to perceive such changes and the effects they have on our freedoms, rights and expectations. If a national government needs to pass through a whole legislative process to approve a new privacy law, an operator like Facebook can change a few lines of code and yield substantial impact on users' privacy profiles. With hundreds of thousands of modifications on platforms like these each year, it is easy to comprehend the reach of this kind of issue. Moreover, many of these changes are temporary, beta versions, running in parallel for different users for A/B testing purposes, making the situation even more complex.

Things get even more radical in the case of algorithmic governance of processes, where technological entities assume progressively higher degrees of agency (and opacity). The Flash Crashes of the stock markets in 2010 are a demonstration: autonomous algorithmic agents gone berserk causing losses for billions of dollars, outside of any legal or cultural or perceptive framework (Menkveld, 2013).

As a result, the levels of power and control exercised on human beings and their societies by the systems that they use are augmenting at exponential levels, and there are progressively fewer and less effective ways for people to perceive and comprehend such processes.

On top of all of this, the dissemination of interfaces ubiquitously across devices, applications, websites and other products and services for which today everything can represent a front-end for digital and data based systems, further augments the incapability to understand the data and information which is captured from our behaviour and its flows and uses (Weber, 2010).

Sharing a picture of our holidays at the beach on social networking sites does not imply the fact that it is clear, for us, that we are producing marketing relevant data about our tastes, consumption levels and geographical locations. And neither is the fact that while using wearable technologies or smart IoT appliances in our daily lives the data that gets captured can be used for marketing, health, insurance, financial and even job purposes.

Furthermore, the rise of the Stacks (Madrigal, 2012) and, more in general, of "walled gardens", or those situations in which applications, services and products pertain to closed, proprietary ecosystems which are not open source and for which both the front-ends and back-ends of the systems are opaque and inaccessible for inspection and understanding further aggravate this problem.

Both those applications directly and, indirectly, the service levels they provide (for example through APIs, social logins, application frameworks) on the one hand make applications and services easy and rapid to develop and deploy, but, on the other hand, subject them to the concentration of power which these large operators represent. It is very convenient to design and develop anything from online services to network-connected physical products using, for example, Google's, Apple's, or Facebook's platforms and services. But, by doing this, it is automatic that our products and services start producing data and information for these large operators, allowing them to interconnect these across a rich variety of domains: if I develop application A and someone else develops application B which is completely different, and we both use, for example, Facebook's social login to implement access services, Facebook will benefit from the data generated from both applications, from the analytics which it desires to capture without even sharing them with A or B, and will be also able to interconnect both data flows with their own. For example, if application A captures, for example, my geographic location (it is, for example, an application which allows me to find where I parked my car) and I have configured my Facebook account so that Facebook is not allowed to know my geographic location, Facebook will have my position anyway, through application A. This kind of reasoning can be applied to all the applications, products and services that use these frameworks.

These facts are valid and relevant for the users of these platforms, but also for the people conceiving and creating these systems, including designers, engineers, managers, administrators, public and private, who progressively lose the possibility (culturally and technically) to understand the implications of their designs.

# 3. Bubbles, Guinea Pigs

An evidence of this occurrence is the emergence of knowledge and information "bubbles".

In the age of Hyperconnectivity (Wellman, 2001) information abundance quickly turns into information overload (O'Reilly, 1980). Therefore, relevance becomes an invaluable competitive advantage and attention a precious currency (Davenport, Beck, 2013).

This is why large operators (from social media services, to search engines, to news and media operators, all the way up to the ones which extract information from devices, appliances and other services) use specific algorithms to try to interpret users' behaviors to try to understand which content might be more relevant for them, filtering out all the rest (or giving it minor priority, visually or hierarchically) and, at the same time, ensuring that the content which generates more revenue for them is granted higher shares of our attention space, to maximize earnings.

These algorithms and software agents also have the effect of tendentially excluding all the rest, closing us in "bubbles", in which what is outside is not even perceived, or very hard to perceive (Pariser, 2011).

Information spectacularization (for example through data and information visualization) further weights down on these processes. Bratton (2008) describes how spectacularized information visualizations (also called "data smog") "distance people—now 'audiences' for data—even further from their abilities and responsibilities to understand relationships between the multiple ecologies in which they live, and the possibilities for action that they have."

These elements – bubbles, algorithmic governance of information and information spectacularization –, thus, may bear the possibility that individuals progressively inhabit a controlled infosphere, in which a limited number of subjects is able to determine what is accessible, usable and, most important of all, knowable.

This power asymmetry also implies the fact that users can systematically be unknowingly exposed to experiments intended to influence their sphere of perception to drive them to adopt certain behaviors over other ones.

This is exactly what happened with Facebook in 2014 (Rushe, 2014; Booth 2014). In an experiment (Kramer et al, 2014), Facebook manipulated information appearing on 689 thousand users' homepages to study the phenomenon of "emotional contagion" answering the question: how to users' emotional expression change when they are exposed to content which is emotionally characterized in specific ways. By algorithmically filtering in or out content with specific characteristics they were able to induce particular expressions. The study (Kramer et al, 2014) concluded: "Emotions expressed by friends, via online social networks, influence our own moods, constituting, to our knowledge, the first experimental evidence for massive-scale emotional contagion via social networks."

This is not the first case: dozens of other experiments (Hill, 2014) dealing with hundreds of thousands of unknowing users included analyses of A/B tests, content filtering for specific purposes, comment and interaction analysis for predictions, spreading of rumors and manufactured information, self-censorship, social influence in advertising, and more.

In 2014, Jonathan Zittrain described an experiment in which Facebook attempted a civic-engineering feat to answer the question: "Could a social network get otherwise-indolent people to cast a ballot in that day's congressional midterm elections?" (Zittrain, 2014). The answer was positive. And the past 2016 elections also demonstrated further ways in which massive, algorithmic controlled social media interactions can influence the determination of major events.

In her article describing the effects of computational agency during the Ferguson protests, Zeynep Tufekci described:

> "Computation is increasingly being used to either directly make, or fundamentally assist, in gatekeeping decisions outside of online platforms. [...] Computational agency is expanding into more and more spheres. Complex, opaque and proprietary algorithms are increasingly being deployed in many areas of life, often to make decisions that are subjective in nature, and hence with no anchors or correct answers to check with. Lack of external anchors in the form of agreed-upon 'right' answers makes their deployment especially fraught. They are armed with our data, and can even divine private information we have not disclosed. They are interactive, act with agency in the world, and are often answerable only to the major corporations that own them. As the internet of things and connected, 'smart' devices become more widespread, the data available to them, and their opportunities to act in the world will only increase. And as more and more corporations deploy them in many processes from healthcare to hiring, their relevance and legal, political and policy importance will also rise." (Tufekci, 2015)

# 4. Interface and Data Biopolitics

The scenario described in the previous sections has important impacts on the "knowability", "readability", accessibility and usability of the world, both in how people use it and interact with it, and in how they are able to design it.

The implications, together with the systematicity and opaqueness of the scenario, calls for the emergence of new areas of scientific, technological and humanistic investigation which can be defined as Interface and Data Biopolitics.

There are multiple definitions for the term "biopolitics": Kjellén's organicist view and his description of the "civil war between social groups" (Lemke, 2011); the political application of bioethics (Hughes, 2004); the interplay between biology and political science (Blank, 2001); Hardt and Negri's (2005) anti-capitalist insurrection through daily life and the body; Foucault's (1997) "biopower", through governments and organizations applying political power to all aspects of human life; and many more.

We refer here mainly to Foucault's definition, which described biopolitics as "a new technology of power...[that] exists at a different level, on a different scale, and [that] has a different bearing area, and makes use of very different instruments". (Foucault, 1997, p. 242)

In his analysis Foucault mainly referred to national states and institutions. Therefore his observations need adaptations to be considered in today's globalized, financial, digital economies and political apparatuses of power. For example the rise of large corporations, which match the power, influence, and reach of national states, the different role of money, its virtualization, and the "finacialization of life" (Lapavitsas, 2013) are things that need to be integrated in such frameworks.

Fundamentally, Biopolitics can be defined as the study of systems as they leverage as many manifestations as possible of our daily lives, activities, relations and bodies to exercise power and control over their users and participants, in explicit and implicit ways.

As demonstrated in the previous sections, today's scenarios of Hyperconnectivity bring about multiple forms of biopolitically relevant contexts. Online and application interfaces, biometrics, wearable computing, IoT, social media and, in general, all human activities with a direct or indirect digital information counterpart generate data which is harvested by large operators in order to be processed to influence our actions, behaviors, beliefs and perceptions, and, thus, to exercise power.

The shift to the digital sphere also provokes a shift from "biopower" to "neuropower" (Väliaho, 2014), as the medium for control shifts from body to mind.

For example, the elements forming an interface exercise a certain degree of power on their users. If only options A and B are available on an interface, the user will not be able to adopt option C. In many cases the user will not even be able to perceive that option C is possible. Hence, the interface, its designer, and the ideology and strategy that comes with both, have a degree of authoritarian agency over the user.

While registering to online services, many times users are asked to select their gender, to characterize their online profile. If, for example, only the "male" and "female" options are available, other options will be excluded and, thus, this could prove to be a problematic, upsetting and troubling scenario for those who feel neither "male" nor "female". The business requirements of the operator, who would need to tag the users with predefined categories that are convenient to be commercialized to marketing and advertising partners, would have the prevalence.

In another scenario, a wearable biometric device could record data for health purposes. For example, a recording of a level of 1.5 to 1.8 from the device for a certain bodily value could indicate a "healthy" condition. If users had a readout of 1.6 they would be considered "healthy", maybe corroborating the fact with a reassuring green light visible on the device, or on the associated application. If, for any plausible reason, the "health" threshold would be changed to a 1.7-2.0 range, the same users would be described as "not healthy". The light would turn to red, maybe accompanied by a message: "visit your doctor!" The body of the users would be the same. They wouldn't feel an additional headache or hurt in some other part of their body. By simple variation of a parameter their status would change, accompanied by a series of authoritative notifications.

This is a very powerful condition. Even taking simpler, less radical and more common examples still shows how a direct possibility to exercise power through asymmetric capacity of capturing, processing and visualizing data, and through designing interfaces in certain ways is available to the operators which own these platforms, systems, devices and services.

With the Internet of Things, this scenario manifests ubiquitously, affecting appliances, our homes through domotics, our schools, offices, stores and, potentially, the public, private and intimate spaces and contexts of our lives. As Pasquinelli (2008) puts it: "it is impossible to destroy the machine, as we ourselves have become the machine."

On top of that, the power asymmetry manifests itself also in another way. While it is users that generate data and information by using interfaces, services and products, at the same time this data is not available to them, nor they have the possibility to perceive the full spectrum of its implications (Blanke et al, 2015).

As of today, most online services offer opportunities for users to download their own data (for example through "Google Takeout"). But these options are misleading, because they let users download their "content", but not the data, information and knowledge that was generated through it by processing it. For example, there is no way for users to know in which marketing categories they have been classified, or what actions they performed led to being classified in such ways.

For example, let's pretend that Facebook identified the category of "potential terrorists" as their machine learning processes discovered a pattern in the frequency with which radical extremists use the letter 'f' in their messages. If certain users, by complete chance, created messages using the same 'f' frequency, they would be classified as "potential terrorists". They would know nothing about it, and this could have implications on their freedoms and rights. Of course this is a paradoxical example, just to make clear the dynamics of this phenomenon.

Moreover, all this data capturing and processing is designed, as stated in the previous section, to confront with relevance and attention, thus resulting in information, knowledge and relations bubbles. While these processes are useful in the scenario of information overload, they also progressively lock out difference from users' perception: the more we are exposed to content which we "potentially like" and to "people we potentially agree with", the more "otherness" disappears from our reach. This brings on a series of negative effects, such as the diminished sensibility to and acceptance of diversity (Bozdag, van den Hoven, 2015), rising levels of cognitive biases (Bozdag, 2013), diminished tolerance, social separation (Ford, 2012), and more.

# 5. Conclusions: Implications for Design

The scenarios described in this article pose great challenges for Designers and, most important of all, for Design Education.

On a first level of inspection, it is simple to verify how all of these situations and configurations of power schemes, practices and behaviors are at the border of what is assessed by laws, regulations, habits and customs. They are at the same time familiar and new, unexpected, unforeseen, unsought. To confront with these issues, approaches which are trans-disciplinary are needed, because no single discipline alone is able to cover all of the knowledge, attitude, perspective which are needed to grasp and understand them.

The possibilities and opportunities to meaningfully deal with the issues presented in the article emerge only at the intersections between law, psychology, culture, philosophy, sociology, ethics and other sciences, humanities and practices.

This fact represents an important opportunity for design, which can act as a convenient, practical and methodologically sound interconnector among disciplines and approaches.

For this, it is of utmost importance that Design curricula natively host such trans-disciplinary approaches, not only combining disciplines as it is common practice in multi-disciplinariety, but traversing them, generating not only contaminations, but also methodological boundary shifts.

The same state of necessity can be detected also for the topics of openness, transparency and access. As seen in the previous sections of the article, most of the times power asymmetries manifest themselves through lack of openness, transparency and access.

Interoperability, data openness and accessibility, usage of open licensing schemes, use of open formats, open access to APIs: these are all types of practices that enable to confront with these problems.

These topics should be standard part of any form of design education, highlighting not only the fact that they enable the emergence of the ethical approaches necessary to resolve the issues described in the article, but, also, represent potential competitive advantages for any organization, as well as the opportunity to create meaningful actions.

The necessity of openness, transparency and access pave the way to another necessary axis for innovation in Design Education, represented by the necessary evolution in which Design needs to confront with Public, Private and Intimate Spheres.

As seen in the previous sections, it is now practically impossible to determine the boundaries of these spheres. Content harvesting, sensors, analytics, and algorithms: these processes know no boundaries. Data and information that appears to be private or even intimate is captured, intercepted, inferred, diverted, producing results for marketing, advertising, or even for surveillance and control. In designing these ecosystems to confront with these issues it is necessary to make every possible effort to clearly and transparently define the boundaries of public, private and intimate spaces, as well as the rights and freedoms which are granted within each of them. This is a complex process, which involves the aforementioned trans-disciplinary approaches as well as considerations that regard current business models, legislations, human rights, and (often national and international) security. There is no simple way to confront with this type of problem. Rather, it is a problem to be dealt with through complex approaches, combining not only different disciplines and practices, but also society as a whole. Here again lies the potential role for design, which can rediscover its humanistic and social elements and act as an interconnector between multiple types of agencies. This is also an evolutionary opportunity for design education practices, in which this modality can be implemented directly into the learning process, by opening it up to the city, the territory and its inhabitants.

Which brings on the next relevant pattern: the one of participation, inclusion and social engagement. Opening up data and processes, using open licenses and formats all are necessary items, but not sufficient. If these actions do not match cultural, social, philosophical ones, they remain ineffective in society. Open Data, as of now, remains a tool for the few, for those researchers, engineers and designers who mediate it for others. For these types of action to become relevant for society design processes must include the patterns of active participation, inclusion and social engagement. This notion must be built into design processes and education, and all possible actions must be performed to inject these ideas into the strategies of those businesses, organizations and, in general, clients who commission the designs.

All leads to the concluding argument of this article, which points out the necessity for design to embrace all possible strategies and actions to promote human dignity, freedom and joy, avoiding atomization and loneliness which have become typical of the years we live in.

The risk society (Beck, 1992) has brought on

> "[...] a mad, Kafkaesque infrastructure of assessments, monitoring, measuring, surveillance and audits, centrally directed and rigidly planned, whose purpose is to reward the winners and punish the losers. It destroys autonomy, enterprise, innovation and loyalty, and breeds frustration, envy and fear. Through a magnificent paradox, it has led to the revival of a grand old Soviet tradition known in Russian as tufta. It means falsification of statistics to meet the diktats of unaccountable power." (Monbiot, 2014)

All this is fundamental to current models that insist on comparison, evaluation and quantification.

Design practice and education can, instead, have a positive role in this, acting as a complex, inclusive and critical interconnector, promoting human dignity, joy and freedom.

About the Authors:

**Salvatore Iaconesi** teaches Near Future Design at ISIA Design Florence. He is a robotic engineer and a philosopher of science. He is TED Fellow, Eisenhower Fellow and Yale World Fellow. He created the Art is Open Source international network, and founded Human Ecosystems.